\documentstyle[11pt]{article}
\newcommand{\D}{{\rm d}}

\begin{document}
\large

 {\Large{\bf
 \centerline {           NEW EXAMPLES}
 \centerline { OF SANDWICH GRAVITATIONAL WAVES}
 \centerline {     AND THEIR IMPULSIVE LIMIT}}}
 \vspace{14mm}
 \centerline { J. Podolsk\' y,  K. Vesel\'y}
 \vspace{5mm}

 {\it
 \centerline {  Department of Theoretical Physics,
           Faculty of Mathematics and Physics,}

 \centerline {   Charles University, V Hole\v sovi\v ck\'ach 2,
     180 00 Prague 8, Czech Republic}}
 \vspace{5mm}

 \centerline{{\small
    Electronic address:  podolsky@mbox.troja.mff.cuni.cz}}
 \vspace{12mm}

\begin{abstract}
Non-standard sandwich gravitational waves are constructed from
the homogeneous $pp$ vacuum solution  and  the motions of free test
particles in the space-times are calculated explicitly.
They demonstrate the caustic property of sandwich waves.
By performing limits to impulsive gravitational wave
it is demonstrated that the resulting particle motions are
identical regardless of the ``initial'' sandwich.

\vspace{3mm}
\noindent
PACS number(s): 04.30.-w, 04.20.Jb, 98.80.Hw
\end{abstract}
\vspace{14mm}

\centerline{\bf 1 Introduction}
\vspace{3mm}

\noindent
Plane-fronted gravitational waves with parallel rays ({\it pp}
waves) are characterized by the existence of a quadruple
Debever-Penrose null vector field which is covariantly constant. In
suitable coordinates (cf. \cite{KSMH}) the metric of vacuum {\it pp}
waves can be written as
\begin{equation}
\D s^2=2\,\D\zeta \D\bar\zeta-2\,\D u\D v-(f+\bar f)\,\D u^2\ , \label{E1}
\end{equation}
where  $f(u,\zeta)$ is an arbitrary function of $u$,
analytic in $\zeta$. The only non-trivial components of the
curvature tensor are proportional to $f_{,\zeta\zeta}$ so that
(\ref{E1}) represents flat Minkowski space-time when $f$ is
linear in $\zeta$. The simplest case for which the metric
describe  gravitational waves arise when $f$ is of the form
\begin{equation}
f(u,\zeta)=d(u)\zeta^2\ , \label{E2}
\end{equation}
where $d(u)$ is an {\it arbitrary} function of $u$;
such solutions are called homogeneous {\it pp} waves
(or ``plane'' gravitational waves).
Performing the transformation (cf. \cite{Penrose})
\begin{eqnarray}
\zeta&=&\frac{1}{\sqrt{2}}\,(Px+iQy)     \ ,\nonumber\\
 v   &=&\frac{1}{2}\,(t+z+PP'x^2+QQ'y^2)  \ ,\label{E3}\\
 u   &=&t-z                               \ ,\nonumber
\end{eqnarray}
where real functions $P(u)\equiv P(t-z)$, $Q(u)\equiv Q(t-z)$
are solutions of differential equations
\begin{equation}
P''+d(u)\,P=0\ ,\qquad Q''-d(u)\,Q=0\ , \label{E4}
\end{equation}
(here prime denotes the derivation with respect to $u$)
the metric can simply be written as
\begin{equation}
\D s^2 = - \D t^2 + P^2 \D x^2 + Q^2 \D y^2 + \D z^2\ . \label{E5}
\end{equation}
This form of the homogeneous {\it pp} waves is suitable for
physical interpretation. Considering two free test particles
standing at fixed $x$, $y$ and $z$, their relative motion in the
$x$-direction is given by the function $P(u)$ while it is given
by $Q(u)$ in the $y$-direction. The motions are unaffected in the
$z$-direction which demonstrate transversality of
gravitational waves. The coordinate $u=t-z$ can now be understood
as a ``retarded time'' and the function $d(u)$ as a ``profile''
of the wave. Note also that functions $P, Q$ may have a higher
degree of smoothness than the function $d$ so that relative
motions of particles are continuous even in  the case of a shock
wave (with a step-function profile, $d(u)\sim\Theta(u)$), or an
impulsive wave (with a distributional profile, $d(u)\sim\delta(u)$).
\vspace{20mm}

\centerline{\bf 2 Standard sandwich wave}
\vspace{3mm}

\noindent
A sandwich gravitational wave \cite{BPR,BP} is constructed from
the homogeneous {\it pp} solution (\ref{E1}), (\ref{E2}) if the function
$d(u)$ is non-vanishing only on some finite interval of $u$, say
$u\in[u_1, u_2]$. In such a case the space-time splits into three
regions: a flat region $u<u_1$ (``Beforezone''), a curved region
$u_1<u<u_2$ (``Wavezone''), and another flat region $u_2<u$
(``Afterzone''). In the region $u<u_1$ where $d(u)=0$ it is natural
to choose  solutions of Eqs. (\ref{E4}) such that $P=1$ and $Q=1$
so that the metric (\ref{E5}) is explicitly written in Minkowski
form. The form of the metric (\ref{E5}) for $u>u_1$ is then given
by solutions of Eqs. (\ref{E4}) where the function $P, Q$ are
chosen to be continuous up to the first derivatives at $u_1$ and
$u_2$.

A standard example of a sandwich wave can be found in textbooks
(cf. \cite{Rindler}). The ``square'' profile function $d(u)$ is
given simply by
\begin{equation}
d(u)=\left\{
  \begin{array}{l} 0, \qquad             u<0     \\
                   a^{-2}, \quad      0\le u\le a^2    \\
                   0, \qquad             a^2<u
  \end{array}\right.
\label{E6}
\end{equation}
where $a$ is a constant. It is easy to show that the
corresponding functions $P, Q$ are given by
\begin{eqnarray}
P(u)&=&\left\{
  \begin{array}{l} 1,    \hskip 57mm             u\le0     \\
                   \cos(u/a), \hskip44mm   0\le u\le a^2         \\
                   -u \sin a/a+\cos a+a\sin a, \hskip11mm  a^2\le u
  \end{array}\right.
\label{E7}\\
Q(u)&=&\left\{
  \begin{array}{l} 1,    \hskip 57mm             u\le 0     \\
                   \cosh(u/a), \hskip42mm 0\le u\le a^2         \\
                   u \sinh a/a+\cosh a-a\sinh a, \qquad a^2\le u
  \end{array}\right.
\label{E8}
\end{eqnarray}
Therefore, particles which were in rest initially accelerate
within the wave in such a way that they approach in
$x$-direction and move apart in  $y$-direction. Behind the wave
they move uniformly (see Fig. 1a).
\vspace{10mm}

\centerline{\bf 3 Non-standard sandwich waves}
\vspace{3mm}

\noindent
Now we construct some other (non-trivial) sandwich waves. Our
work is motivated primarily by the possibility of obtaining
impulsive gravitational waves by performing appropriate limits
starting from {\it different} sandwich waves  (see next Section).
This also enables us to study particle motions in such radiative
space-times. Moreover, the standard sandwich wave given by (\ref{E6})
is very special and ``peculiar'' since it represents a {\it radiative}
space-time containing {\it stationary} regions. Indeed, for $d(u)$
being a positive constant, the Killing vector $\partial_u$
is timelike where $|Re\,\zeta|>|Im\,\zeta|$. This ``strange''
property remained unnoticed in literature so far. It may be useful
to introduce more general sandwich waves which are {\it not} stationary.
\vspace{5mm}

{\bf a) Sandwich wave with  ``$\bigwedge$'' profile}
\vspace{3mm}

Let us consider a solution (\ref{E1}),  (\ref{E2}) for which the
function $d(u)$ takes the form
\begin{equation}
d(u)=\left\{
  \begin{array}{l} 0,           \hskip27mm   u\le -a     \\
                   b\,(a+u)/a,  \hskip10mm  -a\le u\le0         \\
                   b\,(a-u)/a,  \hskip10mm   0\le u\le a         \\
                   0,           \hskip27mm   a\le u
  \end{array}\right.
\label{E9}
\end{equation}
where $a, b$ are arbitrary real (positive) constants.
The wave has a ``wedge'' profile illustrated in Fig. 1b.
Straightforward but somewhat lengthy calculations give the
following form of the functions $P(u), Q(u)$ (continuous up to
the second derivatives everywhere including the points $u=-a$,
$u=0$ and $u=a$):
\begin{eqnarray}
P(u)&=&\left\{
  \begin{array}{l} 1,    \hskip 70mm             u\le -a     \\
  c\,\sqrt{u_1}\,J_{-\frac{1}{3}}(\frac{2}{3}u_1^{3/2}),
                    \hskip40mm -a\le u\le 0         \\
  \sqrt{u_2}\,\left[A\,J_{ \frac{1}{3}}(\frac{2}{3}u_2^{3/2})
                 +B\,J_{-\frac{1}{3}}(\frac{2}{3}u_2^{3/2})\right],
                    \hskip10mm 0\le u\le a         \\
  C\,u+D,           \hskip57mm   a\le u
  \end{array}\right.
\label{E10}\\
Q(u)&=&\left\{
  \begin{array}{l} 1,    \hskip 70mm             u\le -a     \\
  c\sqrt{u_1}\,I_{-\frac{1}{3}}(\frac{2}{3}u_1^{3/2}),
                    \hskip40mm -a\le u\le 0         \\
  \sqrt{u_2}\,\left[E\,I_{ \frac{1}{3}}(\frac{2}{3}u_2^{3/2})
                 +F\,I_{-\frac{1}{3}}(\frac{2}{3}u_2^{3/2})\right],
                    \hskip10mm 0\le u\le a         \\
  G\,u+H,           \hskip57mm   a\le u
  \end{array}\right.
\label{E11}
\end{eqnarray}
where $c=3^{-1/3}\Gamma(\frac{2}{3})$ ($\Gamma$ being the gamma
function), $J_n$ is the
Bessel function, $I_n$ is the modified
Bessel function (cf. \cite{Abram}),
\begin{equation}
u_1=\sqrt[3]{b/a}\,(a+u)\ ,\qquad u_2=\sqrt[3]{b/a}\,(a-u)\ ,\qquad
 \label{E12}
\end{equation}
and $A, B, C, D, E, F, G, H$ are real constants given by
the relations
\begin{eqnarray}
&&A=-2cZ\beta\delta  \ ,\hskip25mm
  B= cZ(\beta\gamma+\alpha\delta)   \ ,\nonumber\\
&&C=-A \sqrt[3]{9b/a}\,/\,\Gamma(1/3)   \ ,\hskip6.3mm
  D= B/c-Ca   \ ,\label{E13}\\
&&E=-2cZ\nu\sigma \ ,\hskip25mm
  F= cZ(\nu\rho+\mu\sigma)  \ ,\nonumber\\
&&G=-E \sqrt[3]{9b/a}\,/\,\Gamma(1/3)   \ ,\hskip6.3mm
  H= F/c-Ga   \ ,\nonumber
\end{eqnarray}
with $Z=\frac{2\pi}{3\sqrt3}(a\sqrt b)^{1/3}$,
\begin{eqnarray}
&&\alpha=J_{\frac{1}{3}}(\kappa)\ ,\hskip27mm
   \beta=J_{-\frac{1}{3}}(\kappa)  \ ,\nonumber\\
&&\gamma= (a\sqrt b)^{2/3}\,J_{-\frac{2}{3}}(\kappa)\ ,\hskip9.5mm
  \delta=-(a\sqrt b)^{2/3}\,J_{\frac{2}{3}}(\kappa)   \ ,\label{E14}\\
&&\mu=I_{\frac{1}{3}}(\kappa)   \ ,\hskip28mm
  \nu=I_{-\frac{1}{3}}(\kappa)  \ ,\nonumber\\
&&\rho =(a\sqrt b)^{2/3}\,I_{-\frac{2}{3}}(\kappa)  \ ,\hskip10mm
 \sigma=(a\sqrt b)^{2/3}\,I_{\frac{2}{3}}(\kappa)   \ ,\nonumber
\end{eqnarray}
$\kappa=\frac{2}{3}a\sqrt b$
(note that $\beta\gamma-\alpha\delta=1/Z=\nu\rho-\mu\sigma$).
Typical behaviour of the particles in the above sandwich
space-time is shown in Fig. 1b.
\newpage

{\bf b) Sandwich wave with  ``$/\!\!\hskip1pt|$'' profile}
\vspace{3mm}

Another sandwich wave can be obtained using the  function $d(u)$ such that
\begin{equation}
d(u)=\left\{
  \begin{array}{l} 0,           \hskip27mm   u\le -a     \\
                   b\,(a+u)/a,  \hskip10mm  -a\le u<0         \\
                   0,           \hskip27mm   0<u
  \end{array}\right.
\label{E15}
\end{equation}
where $a, b$ are again constants. In fact, it has a ``saw'' profile
(see Fig. 1c) which is one ``half'' of the sandwich discussed
above. It is non-symmetric and contains two discontinuities of different
types. The functions $d(u)$ defined by Eq. (\ref{E9}) and
(\ref{E15}) coincides for $u\le 0$ so that the functions $P, Q$
are identical in both cases. It is only necessary to join the
solution at $u=0$ differently:
\begin{eqnarray}
P(u)&=&\left\{
  \begin{array}{l} 1,    \hskip 70mm             u\le -a     \\
  c\,\sqrt{u_1}\,J_{-\frac{1}{3}}(\frac{2}{3}u_1^{3/2}),
                    \hskip40mm -a\le u\le 0         \\
  K\,u+L,           \hskip57mm  0\le u
  \end{array}\right.
\label{E16}\\
Q(u)&=&\left\{
  \begin{array}{l} 1,    \hskip 70mm             u\le -a     \\
  c\,\sqrt{u_1}\,I_{-\frac{1}{3}}(\frac{2}{3}u_1^{3/2}),
                    \hskip40mm -a\le u\le 0         \\
  M\,u+N,           \hskip57mm  0\le u
  \end{array}\right.
\label{E17}
\end{eqnarray}
where
\begin{equation}
K = c\delta\,\sqrt[3]{b/a}\ ,\qquad
L = c\beta\,\sqrt[6]{a^2 b}\ ,\qquad
M = c\sigma\,\sqrt[3]{b/a}\ ,\qquad
N = c\nu\,\sqrt[6]{a^2 b}  \ .\label{E18}
\end{equation}
Relative motions of test particles are illustrated in Fig. 1c.
Qualitatively, they resemble motions in both  previous cases
(cf. Fig. 1a and Fig. 1b), only the relative velocities of
particles in the region behind the wave
(given in  $x$-direction by $\ -\sin a/a$, $C$, and $K$ ,
respectively; in  $y$-direction by $\ \sinh a/a$, $G$, and $M$)
depend differently on particular choice of the parameters $a$ and $b$.
\vspace{3mm}

{\bf c) Asymptotic sandwich wave}
\vspace{3mm}

Let us also consider the  function $d(u)$ of the form
\begin{equation}
d(u)=\frac{n}{2} \exp(-n|u|)   \ ,
\label{E19}
\end{equation}
(shown in Fig. 1d) where $n$ is an arbitrary real positive constant .
Now there are {\it no flat regions} in front of the wave and
behind it. The space-time is curved everywhere (it is of Petrov
type~N and therefore radiative), it becomes flat only
asymptotically as $u\to\pm\infty$. For this reason we choose the
functions $P, Q$ such that $P(u\to-\infty)\to 1$,
$P'(u\to-\infty)\to 0$ and similarly $Q(u\to-\infty)\to 1$,
$Q'(u\to-\infty)\to 0$. Then it can be shown that the functions
are given by
\begin{eqnarray}
P(u)&=&\left\{
  \begin{array}{l}
  J_0\left(\sqrt{\frac{2}{n}}\exp(\frac{n}{2}u)\right),
                    \hskip63mm  u\le 0         \\
    A_1\, J_0\left(\sqrt{\frac{2}{n}}\exp(-\frac{n}{2}u)\right)
   +A_2\, Y_0\left(\sqrt{\frac{2}{n}}\exp(-\frac{n}{2}u)\right),
           \hskip10mm  0\le u
  \end{array}\right.
\label{E20}\\
Q(u)&=&\left\{
  \begin{array}{l}
  I_0\left(\sqrt{\frac{2}{n}}\exp(\frac{n}{2}u)\right),
                    \hskip63.5mm  u\le 0         \\
    B_1\, I_0\left(\sqrt{\frac{2}{n}}\exp(-\frac{n}{2}u)\right)
   +B_2\, K_0\left(\sqrt{\frac{2}{n}}\exp(-\frac{n}{2}u)\right),
           \hskip10mm  0\le u
  \end{array}\right.
\label{E21}
\end{eqnarray}
where
\begin{eqnarray}
&&A_1=-\frac{\pi}{2}\lambda(\tilde\alpha\tilde\delta+\tilde\beta\tilde\gamma)
  \ ,\hskip15.0mm
  A_2=\pi\lambda\tilde\alpha\tilde\gamma  \ ,\label{E22}\\
&&B_1=\lambda(\tilde\mu\tilde\sigma+\tilde\nu\tilde\rho)
  \ ,\hskip21.5mm
  B_2=2\lambda\tilde\mu\tilde\rho
    \ ,\nonumber
\end{eqnarray}
with
\begin{eqnarray}
&&\tilde\alpha=J_0(\lambda)\ ,\hskip8mm
  \tilde\beta =Y_0(\lambda)  \ ,\hskip8mm
  \tilde\gamma=J_1(\lambda)\ ,\hskip8mm
  \tilde\delta=Y_1(\lambda)   \ ,\label{E23}\\
&&\tilde\mu=I_0(\lambda)   \ ,\hskip9mm
  \tilde\nu=-K_0(\lambda)  \ ,\hskip4mm
\tilde\rho =I_1(\lambda)  \ ,\hskip8mm
  \tilde\sigma=K_1(\lambda)   \ ,\nonumber
\end{eqnarray}
$\lambda=\sqrt{2/n}$ (note also that
$\tilde\alpha\tilde\delta-\tilde\beta\tilde\gamma=
-\frac{2}{\pi\lambda}$ and
$\tilde\mu\tilde\sigma-\tilde\nu\tilde\rho=-\frac{1}{\lambda}$).
Typical behaviour of the functions $P, Q$ is shown in Fig. 1d. It can
be observed that in both asymptotic regions $u\to\pm\infty$
the particles move uniformly.

\vspace{10mm}

\centerline{\bf 4 Impulsive limit}
\vspace{3mm}
Now we can use the above results to construct impulsive gravitational
waves. For standard sandwich wave (\ref{E6})-(\ref{E8}) it is easy to perform
the limit $a\to0$. Then the profile function $d(u)$ approaches
the $\delta$ function distribution and, using $\sin a/a\to 1$,
$\sinh a/a\to 1$ we get
\begin{eqnarray}
P(u)&=&1-u\,\Theta(u)\ ,\nonumber\\
Q(u)&=&1+u\,\Theta(u)\ ,\label{E24}
\end{eqnarray}
where $\Theta$ is the Heaviside step function ($\Theta=0$ for
$u<0$, $\Theta=1$ for $u>0$).

For non-standard sandwiches introduced in previous section one
has to perform similar limits more carefully. It is well known
that the sequence of ``$\wedge$'' functions given by (\ref{E9})
approach the $\delta$ function (in a distributional sense) as
$a\to 0$ if the second parameter is $b=1/a$ (so that the
normalization condition $\int_{-\infty}^{+\infty} d(u)\,\D u=1$
holds for arbitrary $a$). Considering this limit,
$\kappa=\frac{2}{3}\sqrt{a}$, the parameters (\ref{E14}) are
$\alpha\sim\mu \sim a^{1/6}\,  3^{-1/3} /\Gamma(4/3)$,
$\beta \sim\nu \sim a^{-1/6}\, 3^{1/3}  /\Gamma(2/3)$,
$\gamma\sim\rho\sim 3^{2/3}/\Gamma(1/3)$,
$\delta\sim-\sigma\sim-a^{2/3}\,3^{1/3}/(2\Gamma(2/3))$,
and (\ref{E13}) gives
\begin{equation}
C\to-1 \ ,\quad  D\to 1 \ ,\quad
G\to 1 \ ,\quad  H\to 1 \ .\label{E25}
\end{equation}
Therefore, the functions $P, Q$ describing relative motions of test
particles in the corresponding impulsive wave are again given by
(\ref{E24}).

Analogously, the limit $a\to 0$ of ``$/\!\!\hskip1pt|$'' sandwiches given by
(\ref{E15}) with  $b=2/a$ gives
$\beta \sim\nu\sim (2a)^{-1/6}\, 3^{1/3}  /\Gamma(2/3) $,
$\delta\sim-\sigma\sim -(2a)^{2/3}\, 3^{1/3}/(2\Gamma(2/3))$,
so that the parameters (\ref{E18}) are
\begin{equation}
K\to-1 \ ,\quad  L\to 1 \ ,\quad
M\to 1 \ ,\quad  N\to 1 \ .\label{E26}
\end{equation}
Again, the resulting functions $P, Q$ can be written in the form
(\ref{E24}).

Finally, we can perform a limit $n\to\infty$ of ``asymptotic
sandwich waves'' given by  profile functions (\ref{E19}).
For $u\le0$ it follows immediately from (\ref{E20}), (\ref{E21}) that
$P\to1$ and $Q\to1$. For $u\ge0$ calculations are more complicated: for $n\to\infty$
we get
$\tilde\alpha\sim\tilde\mu\sim 1$,
$\tilde\beta \sim -\frac{1}{\pi}\ln n$,
$\tilde\gamma\sim\tilde\rho\sim (2n)^{-1/2}$,
$\tilde\delta\sim -\frac{1}{\pi}[(2n)^{-1/2}\ln n+\sqrt{2n}\,]$,
$\tilde\nu\sim    -\frac{1}{2}\ln n$,
$\tilde\sigma\sim \frac{1}{2} [\sqrt{2n}-(2n)^{-1/2}\ln n\,]$,
so that
$A_1\sim B_1\sim 1$,
$A_2\sim \pi/n$ and $B_2\sim 2/n$. Since
$J_0\sim I_0\sim1 $,
$Y_0\sim-\frac{1}{\pi}[n u+\ln n\,]$ and
$K_0\sim \frac{1}{2}[n u+\ln n\,]$
as $n\to\infty$, in the limit we obtain
\begin{equation}
P\sim  1-u\ ,\qquad\qquad  Q\sim 1+u \ ,\label{E27}
\end{equation}
i.e. the relations (\ref{E24}) are revealed once more.

Note finally that using the transformation (\ref{E3}) with the
functions $P$ and $Q$ given by (\ref{E24}) the metric of the
impulsive homogeneous {\it pp} wave
\begin{equation}
\D s^2 = - \D t^2 + (1-u\Theta(u))^2 \D x^2 +
        (1+u\Theta(u))^2 \D y^2 + \D z^2\ , \label{E28}
\end{equation}
goes over to
\begin{equation}
\D s^2=2\,\D\zeta \D\bar\zeta-2\,\D u\D v
       -\delta(u)(\zeta^2+\bar \zeta^2)\,\D u^2\ , \label{E29}
\end{equation}
i.e. the metric (\ref{E1}) with $f(u,\zeta)=\delta(u)\zeta^2$.
Although this form of the impulsive wave is illustrative with the
pulse evidently localized along the hyperplane $u=0$, the metric
(\ref{E28}) is more convenient in the sense that the metric
system is continuous, $\delta$ function appearing only in the
components of the curvature tensor.

The transformation (\ref{E3}) with (\ref{E24}) also relates to
the ``scissors-and-paste'' approach to the construction of impulsive
solutions \cite{Penrose} which recently enabled  new impulsive
gravitational waves of somewhat different type \cite{Nutku,Hogan}
to be found.
\vspace{10mm}

\centerline{\bf 5 Concluding remarks}
\vspace{3mm}

\noindent
We constructed three new types of non-standard sandwich {\it pp}
waves with ``wedge'' (\ref{E9}), ``saw'' (\ref{E15}) and ``asymptotic''
(\ref{E19}) profiles. Contrary to the standard sandwich wave
(\ref{E6}) they do not contain stationary regions. Particle
motions were calculated explicitly and the corresponding limits
to impulsive gravitational wave were performed. It was
shown that all these limits give the same result (\ref{E24}).
Moreover, for sandwich waves presented above there exist critical
values of $u$ for which the function $P(u)$ vanishes so that all
the particles initially at rest on the $x$-axis collide. This
demonstrates the caustic property of (plane) sandwich waves
\cite{BP}.
\vspace{5mm}

\centerline{\bf  Acknowledgments}
\vspace{3mm}

\noindent
We acknowledge the support of grants No. GACR-202/96/0206 and
No. GAUK-230/1996 from the Czech Republic and Charles University.

\newpage
{\LARGE Figure Captions}

\bigskip
\bigskip
\noindent
Fig. 1. Typical exact behaviour of functions $P(u)$ and $Q(u)$
determining relative motions of free test particles (initially at
rest) in $x$ and $y$-directions, respectively, caused by sandwich
gravitational waves of various profiles: a) standard ``square'',
b) ``wedge'', c) ``saw'', d) ``asymptotic'' sandwich wave. In the last case
also the limiting procedure $n\to\infty$ leading to an impulsive
wave is indicated by corresponding dashed lines.

\end{document}